\documentclass[apjl]{emulateapj}
\usepackage{fancyhdr}
\usepackage{color}
\usepackage{natbib}
\usepackage{amssymb,amsmath}
\usepackage{graphicx}
\usepackage{multirow}
\shorttitle{Gemini Planet Imager First $H$-Band Spectrum of Beta Pictoris \MakeLowercase{b}}
\shortauthors{Chilcote et al.}
\begin{document}
\bibliographystyle{apj}

\submitted{Submitted to the Astrophysical Journal Letters, 2014 July 15}
\title{The First H-band Spectrum of the Massive Gas Giant Planet Beta Pictoris \MakeLowercase{b}

with the Gemini Planet Imager}

%% ------------------------- AUTHORS ------------------------- %%
\author{Jeffrey Chilcote\altaffilmark{1}, Travis Barman\altaffilmark{2}, Michael P. Fitzgerald\altaffilmark{1}, James R. Graham\altaffilmark{3}, James E. Larkin\altaffilmark{1}, Bruce Macintosh\altaffilmark{4,5}, Brian Bauman\altaffilmark{5}, Adam S. Burrows\altaffilmark{6}, Andrew Cardwell\altaffilmark{7}, Robert J. De Rosa\altaffilmark{8}, Daren Dillon\altaffilmark{9}, Rene Doyon\altaffilmark{10}, Jennifer Dunn\altaffilmark{11}, Darren Erikson\altaffilmark{11}, Donald Gavel\altaffilmark{9}, Stephen J. Goodsell\altaffilmark{12}, Markus Hartung\altaffilmark{7}, Pascale Hibon\altaffilmark{7}, Patrick Ingraham\altaffilmark{4}, Paul Kalas\altaffilmark{3}, Quinn Konopacky\altaffilmark{13}, J\'{e}r\^{o}me Maire\altaffilmark{13}, Franck Marchis\altaffilmark{14}, Mark S. Marley\altaffilmark{15}, Christian Marois\altaffilmark{11}, Max Millar-Blanchaer\altaffilmark{13}, Katie Morzinski\altaffilmark{16}, Andrew Norton\altaffilmark{9}, B. R. Oppenheimer\altaffilmark{17}, David Palmer\altaffilmark{5}, Jennifer Patience\altaffilmark{8}, Marshall D. Perrin\altaffilmark{18}, Lisa Poyneer\altaffilmark{5}, Laurent Pueyo\altaffilmark{18}, Fredrik Rantakyr\"{o}\altaffilmark{7}, Naru Sadakuni\altaffilmark{7}, Leslie Saddlemyer\altaffilmark{11}, Dmitry Savransky\altaffilmark{19}, Andrew Serio\altaffilmark{7}, Anand Sivaramakrishnan\altaffilmark{17,18}, Inseok Song\altaffilmark{20}, Remi Soummer\altaffilmark{18}, Sandrine Thomas\altaffilmark{15}, J. Kent Wallace\altaffilmark{21}, Sloane J. Wiktorowicz\altaffilmark{22,23}, and Schuyler Wolff\altaffilmark{23}}

\affil{$^{1}$ Department of Physics and Astronomy, University of California, Los Angeles, CA USA 90095}
\affil{$^{2}$ Lunar and Planetary Laboratory, University of Arizona, Tucson AZ 85721}
\affil{$^{3}$ Astronomy Department, University of California, Berkeley; Berkeley CA 94720}
\affil{$^{4}$ Kavli Institute for Particle Astrophysics and Cosmology, Stanford University, Stanford, CA 94305}
\affil{$^{5}$ Lawrence Livermore National Laboratory, Livermore, CA USA 94551}
\affil{$^{6}$ Department of Astrophysical Sciences, Princeton University, Princeton, NJ 08544}
\affil{$^{7}$ Gemini Observatory, Casilla 603, La Serena, Chile}
\affil{$^{8}$ School of Earth and Space Exploration, Arizona State University, PO Box 871404, Tempe, AZ 85287, USA}
\affil{$^{9}$ University of California Observatories/Lick Observatory, University of California, Santa Cruz;  Santa Cruz, CA 95064}
\affil{$^{10}$ Observatoire du Mont-M'{e}gantic and D'{e}partement de physique Universit'{e} de Montr'{e}al, Montr'{e}al, QC H3T 1J4, Canada}
\affil{$^{11}$ NRC Herzberg Astronomy and Astrophysics, 5071 West Saanich Rd, Victoria, BC, Canada, V9E 2E7}
\affil{$^{12}$ Gemini Observatory, 670 N. A'ohoku Place, Hilo, HI 96720, USA}
\affil{$^{13}$ Dunlap Institute for Astronomy and Astrophysics University of Toronto, Toronto, Ontario, Canada M5S 3H4}
\affil{$^{14}$ SETI Institute, Carl Sagan Center, 189 Bernardo Avenue,  Mountain View, CA 94043, USA}
\affil{$^{15}$ NASA Ames Research Center,  Mountain View, CA 94035}
\affil{$^{16}$ Steward Observatory, University of Arizona, Tucson AZ 85721}
\affil{$^{17}$ Department of Astrophysics, American Museum of Natural History, New York, NY 10024}
\affil{$^{18}$ Space Telescope Science Institute, Baltimore, MD 21218}
\affil{$^{19}$ Sibley School of Mechanical and Aerospace Engineering, Cornell University, Ithaca, NY 14853}
\affil{$^{20}$ Department of Physics and Astronomy, University of Georgia, Athens, GA 30602}
\affil{$^{21}$ Jet Propulsion Laboratory, California Institute of Technology Pasadena CA 91125}
\affil{$^{22}$ Department of Astronomy, UC Santa Cruz, 1156 High Street, Santa Cruz, CA 95064}
\affil{$^{23}$ NASA Sagan Fellow}
\affil{$^{24}$ Department of Physics and Astronomy, Johns Hopkins University, Baltimore, MD 21218}

%% ------------------------- END AUTHORS ------------------------- %%
%% ------------------------- ABSTRACT ------------------------- %%

\begin{abstract}
Using the recently installed Gemini Planet Imager (GPI), we have taken the first \textit{H}-band spectrum of the planetary companion to the nearby young star beta Pictoris. GPI is designed to image and provide low-resolution spectra of Jupiter sized, self-luminous planetary companions around young nearby stars. These observations were taken covering the \textit{H}-band (1.65 microns). The spectrum has a resolving power of $\sim$ 45 and demonstrates the distinctive triangular shape of a cool substellar object with low surface gravity. Using atmospheric models, we find an effective temperature of 1650 $\pm 50$~K and a surface gravity of $\log(g) = 4.0 \pm 0.25$ (cgs units). These values agree well with predictions from planetary evolution models for a gas giant with mass between 10 and 12 $M_{\rm Jup}$ and age between 10 and 20 Myrs.
\end{abstract}

%% ------------------------- END ABSTRACT ------------------------- %%


%% ------------------------- MAIN TEXT ------------------------- %%

\section{Introduction}

For over a decade, there have been ongoing efforts to directly image young Jupiter mass exoplanets still luminous in the infrared (IR) from their formation process. Examples of such planets include 2M1207b \citep{Chauvin2005}, Fomalhaut b \citep{Kalas2008}, the HR8799 system \citep{Marois2008,Marois2010}, $\beta$ Pic b \citep{Lagrange2010}, IRXS J1609 b \citep{Lafreniere2010}, HD 95086 b \citep{Rameau2013}, and GJ 504 b \citep{Kuzuhara2013}.

Beta Pictoris (HD 39060) is an A6V star located $19.44\pm0.05$ pc from Earth \citep{Gray2006, vanLeeuwen2007}. \citet{Zuckerman2001} estimates the age of \mbox{$\beta$ Pic} at $12_{-4}^{+8}$ Myr, but that has recently been revised upwards to $21\pm4$ Myr  (Binks \& Jefferies 2014)\nocite{BinksJeffries2014}. \mbox{$\beta$ Pic} represents the earliest examples of using high contrast imaging to directly detect a circumstellar disk (Smith \& Terrile 1984)\nocite{SmithTerrile1984}.  The disk is seen edge-on and shows asymmetric structure that has been attributed to planetary perturbations  \citep{Burrows1995,Kalas1995,Golimowski2006,Mouillet1997,Heap2000}.
The planet 
possibly 
responsible for these perturbations was eventually discovered by direct imaging \citep{Lagrange2010}.
$\beta$ Pic b has been detected by VLT/NACO \citep{Lagrange2010}, Gemini/NICI \citep{Boccaletti2013}, Magellan AO \citep{Males2014,Morzinski2014}, and Gemini/GPI \citep{Macintosh2014}. This has led to a multi-epoch attempt to understand the planet's orbital parameters and to discern if it is aligned with the main disk or the secondary inclined disk \citep{Lagrange2012,Chauvin2012,Macintosh2014}.  The basic  properties of $\beta$ Pic b have been estimated using SED fitting of broad band photometry, resulting in an effective temperature of $1700 \pm 100K$, with a \mbox{log $g$ $ = 4.0 \pm 0.5$}\citep{Bonnefoy2013}.  Previous comparisons of the planet's bolometric luminosity and system age to evolutionary cooling tracks resulted in a mass from 9 to 13 M$_\mathrm{Jup}$ \citep{Bonnefoy2013, Males2014}. Using a cross-correlation technique and high-spectral resolution over a narrow wavelength range, \citet{Snellen2014} were able to measure the planet's spin ($vsin(i) \sim 25$ km/s ) and detect carbon monoxide absorption in the $K$ band.

Understanding the atmospheres of these very young giant exoplanets is a challenging task
because we have only a handful of objects to study spectroscopically. 
The theoretical models used to compute the emergent flux from 
these 
planetary 
atmospheres are often extensions of those generated for brown dwarfs, 
yet the spectra of the HR8799 planets exhibit 
significant differences relative to 
brown dwarfs \citep{Barman2011a,Marley2012}. 
Spectroscopy of $\beta$ Pic b offers another opportunity to study the atmospheric properties of a young giant planet that is substantially hotter than the HR8799 planets.

Here we present the first $H$-band spectral mode observations of $\beta$ Pic b
with GPI. An analysis of the orbital parameters using astrometric measurements from
these data has been published in \citet{Macintosh2014}. In section \ref{sec:GPI}, we briefly review the recently delivered Gemini Planet Imager being commissioned on the Gemini South telescope. In section \ref{sec:obs_data_reduction}, we discuss the observations and data reduction used to analyze the spectrum with this new instrument. Analysis of the $H$-band spectrum, along with existing photometry, is presented in section  \ref{sec:results}. Conclusions are discussed in Section \ref{sec:Discussion}.

\section{Gemini Planet Imager}\label{sec:GPI}

The Gemini Planet Imager is a facility class instrument that was designed and built to directly image and spectroscopically characterize young, Jupiter sized, self-luminous extrasolar planets. GPI was built for the Gemini Observatory, and installed at Gemini South in the fall of 2013. The high dynamic ranges involved in directly imaging extrasolar planets required GPI to be designed to pay special attention to speckle suppression \citep{Macintosh2006,Graham2007}. 

GPI uses different sub-systems to combine several key technologies into one instrument. The GPI 
adaptive optics (AO) system incorporates a large number of degrees of freedom and uses a spatially filtered wavefront sensor to enhance contrast near the star.
GPI first light and commissioning tests demonstrate that the AO system lowers the total wavefront error from dynamic sources and quasi-static errors by an order of magnitude compared to earlier AO systems \citep{Macintosh2014}. The GPI AO system is composed of a low spatial frequency, high stroke, 11 actuator diameter woofer deformable mirror, and a 
$64 \times 64$ 
Micro-electro-mechanical-system low stroke, high frequency, deformable mirror from Boston Micromachines  \citep{Poyneer2011}, with a 43-actuator-diameter region for high order corrections. Light travels through a spatially-filtered wave-front sensor, to remove high spatial
frequency signals that would 
violate the sampling theorem 
and be aliased 
as low-frequency signals. Spatial filtering is implemented as a hard-edged 
stop in the focal plane before the wave front sensor \citep{poyneer2003}. 

Diffraction is suppressed by an apodized-pupil Lyot coronagraph \citep{Soummer2011,Macintosh2014}. A grid of narrow, widely-spaced lines printed onto the apodizer forms a two-dimensional grating, producing diffracted images of the central star in a square pattern. These four satallite spots allow for a sampling of the central star spectrum, instrumental, and atmospheric effects in the same image as the object of interest \citep{wang2014}. 

A infrared (IR) calibration wavefront sensor was designed to suppress non-common path wavefront errors \citep{Wallace2010} by providing feedback about these errors to the AO sytem. Finally, the science instrument is a near-IR (1-2.5 $\mu$m) integral field spectrograph (IFS) with an imaging polarimetry mode \citep{Chilcote2012,Larkin2014}. 
The spatial field is sampled by a lenslet array and then dispersed, resulting in \mbox{$\sim$ 37,000} individual spectra with a spectral resolving power R$=\lambda / \delta\lambda \sim 30-90$. The spatial plane is sampled at $14.14\pm0.01$ milliarcseconds per pixel \citep{Konopacky2014}. In first light observations, GPI achieved a 5-$\sigma$ contrast of $10^{5}$ at 0.35 arcseconds and $10^{6}$ at 0.75 arcseconds \citep{Macintosh2014}.

\section{Observations and Data Reduction}\label{sec:obs_data_reduction}

$\beta$ Pic was observed with GPI in the $H$ band 
(1.5072$\mu$m - 1.7899$\mu$m\footnote{Defined by the 80\% power-point of the filters}, $R\sim$44-49) by the GPI Verification and 
Commissioning team on Gemini South during first light and then during the first verification and commissioning runs on 18 November 2013 and 10 December 2013, respectively. During the November observations, 32 individual 59.6-second images were obtained in coronagraphic mode, with the cryocoolers \citep{Chilcote2012,Larkin2014} operating at a reduced power level to reduce the effects of vibration introduced into the telescope. Seeing conditions were on average 0.68\arcsec as measured by the Gemini South DIMM.
As the observations were performed during instrument commissioning, various operational modes were used 
during a specific data set to evaluate performance of the instrument. During the December 2013 observations, 14 individual 59.6-second images were obtained in coronagraphic mode. For eight of the images, the IFS cryocoolers were operating at full power, while in the remaining six images, the cryocoolers were operating in a reduced power state similar to the November observations. Each image has a different spatial filter size \& woofer integrator memory value in an attempt to optimize AO performance \citep{Macintosh2014}. Immediately after the observing sequence was completed, and at the same telescope orientation and flexure, a single observation of the flood illuminated argon calibration source was taken to accurately track the shift of the spectral solution on the 
HAWAII-2RG detector.

The images were first processed using the GPI data reduction pipeline \citep{Perrin2014}. The pipeline requires the location and spectral solution for every lenslet. These lenslet locations were determined by using a cross correlation of the single argon image taken during the observing sequence as $\beta$ Pic and high S/N, deep images made during daytime calibrations. 
The telescope elevation differed between the science images and the daytime calibration sequence. 
The resultant shift was used to determine the overall change of the wavelength solution  
between the daytime calibrations and that appropriate for 
the observations of $\beta$ Pic.

With a shifted wavelength calibration, the GPI data reduction pipeline was used to reduce all images, apply dark corrections, 
remove bad pixels, track satellite spot locations, and convert each microspectra into a 37-channel spectral cube ($1.490-1.802\mu$m). Each data set was processed in an identical way.

Further data processing was done outside of the GPI pipeline. The GPI atmospheric dispersion corrector was not commissioned at the time these observations were made; therefore, each image and each spectral slice are independently registered using the stellar position found by the four satellite spots. GPI is mounted on a Cassegrain port with derotator disabled so each image has a different sky orientation. In post processing, these images are rotated so that the planet has a fixed location. 

Since the satellite spots are imaged at an identical time under identical conditions, in theory their PSFs should closely match the planet PSF especially when the four spots are averaged together. Instrumental effects and atmospheric effects are estimated from satellite spot spectra. An 8000K, log $g$ $=4.0$ BT-Nexgen model \citep{Allard2012} 
convolved to the resolution of GPI, 
was used to approximate the A6V stellar spectrum of $\beta$ Pic A. 
This allows the instrumental and telluric features 
under identical conditions to be estimated for the planet spectrum and removed. 

\begin{figure}
\epsscale{.90}
\plottwo{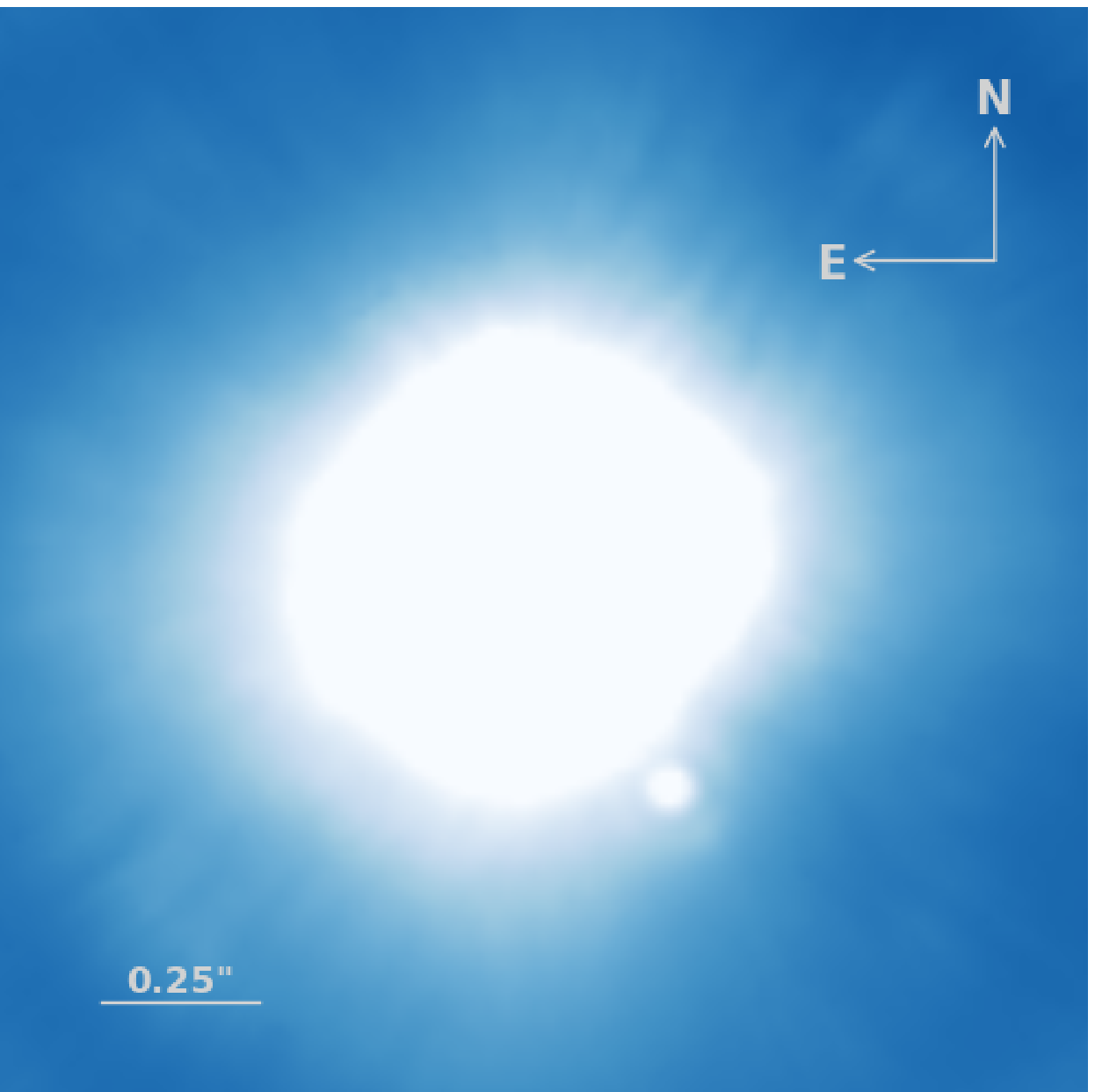}{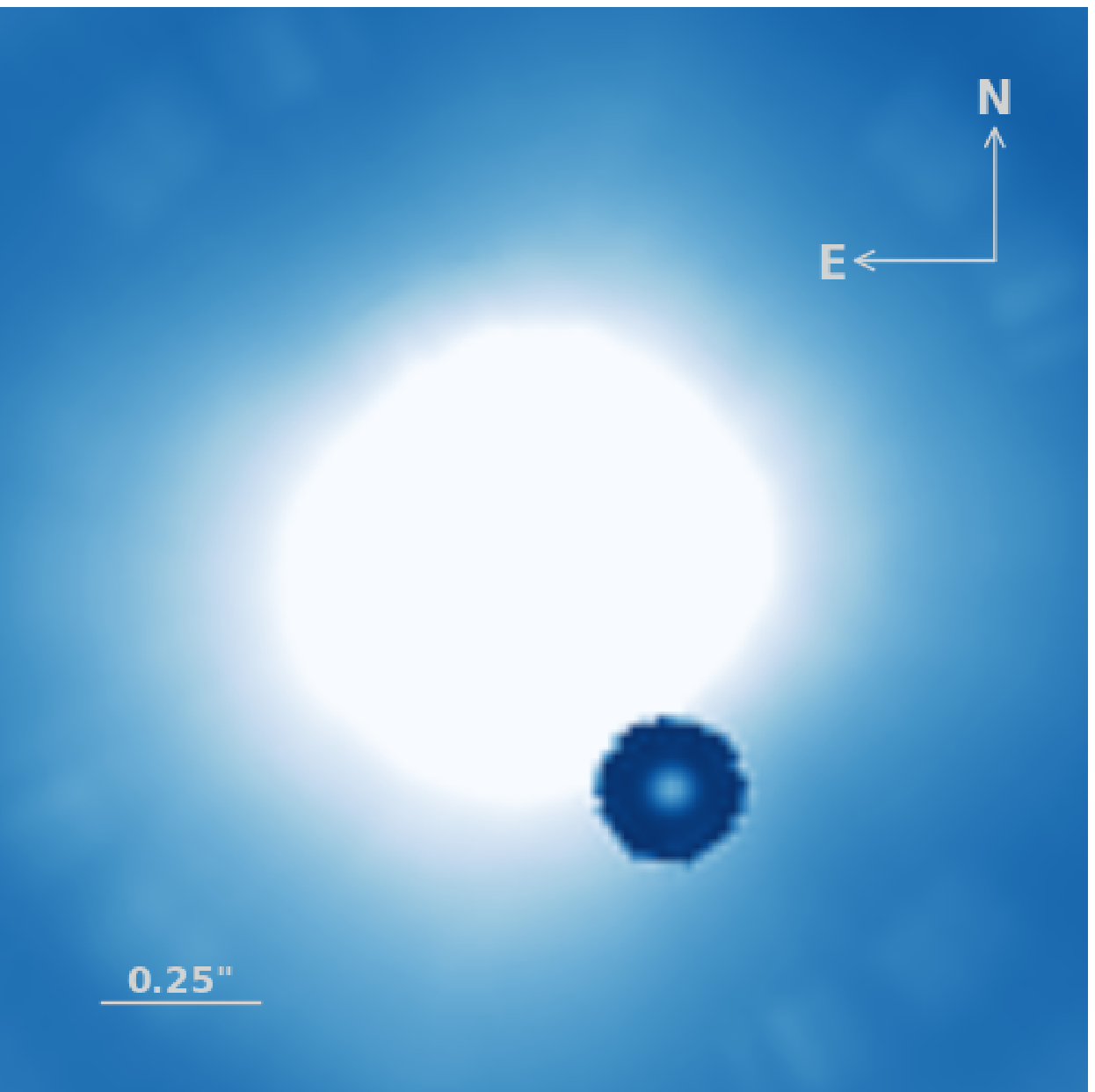}
\caption{{\em Left:} An average combined set of GPI images of
$\beta$ Pic b
from November 2013 with no additional post processing removal of the background. {\em Right:} An average combined set of images from November with a circular annulus defined around the estimated location of the planet, which has been used to define a surface in each image and spectral channel to subtract the remaining halo light. In order to remove this halo, we fit a third-order spline surface to an aperture of radius$=$57.2--114.4 mas centered on the location of the planet, which includes the space around the planet but does not include the planet itself. A PSF, generated by the average of the four satellite spot cores, was scaled and subtracted from the planet position in parallel to the spline fit. Images are averaged along the 37 spectral channels in \textit{H}-band ($\sim1.5-1.8\mu$m).}\label{fig1_2}
\end{figure}

We found that the remaining halo in these initial first light images was smooth, and 
dominated by uncorrected atmospheric halo speckles, rather than quasistatic speckles. 
In order to remove this halo, we fit a third-order spline surface to an aperture of 
radius$=$57.2--114.4 mas centered on the location of the planet, which includes the 
space around the planet but does not include the planet itself. A PSF, generated by 
the average of the four satellite spot cores, was scaled and subtracted from the 
planet position in parallel to the spline fit. This average PSF of the four satellite spots was generated 
for each particular image and wavelength channel to which it corresponds This spline \mbox{surface $+$ reference}
PSF is generated to subtract the smooth halo 
and estimate the flux of the PSF. A Levenberg-Marquardt least-squares minimization 
\citep{2009ASPC..411..251M} was performed to find the best fit of the underlying 
halo and the planet PSF in each image and at each spectral channel (Figure \ref{fig1_2}). 

\begin{figure}
\epsscale{1.0}
\plotone{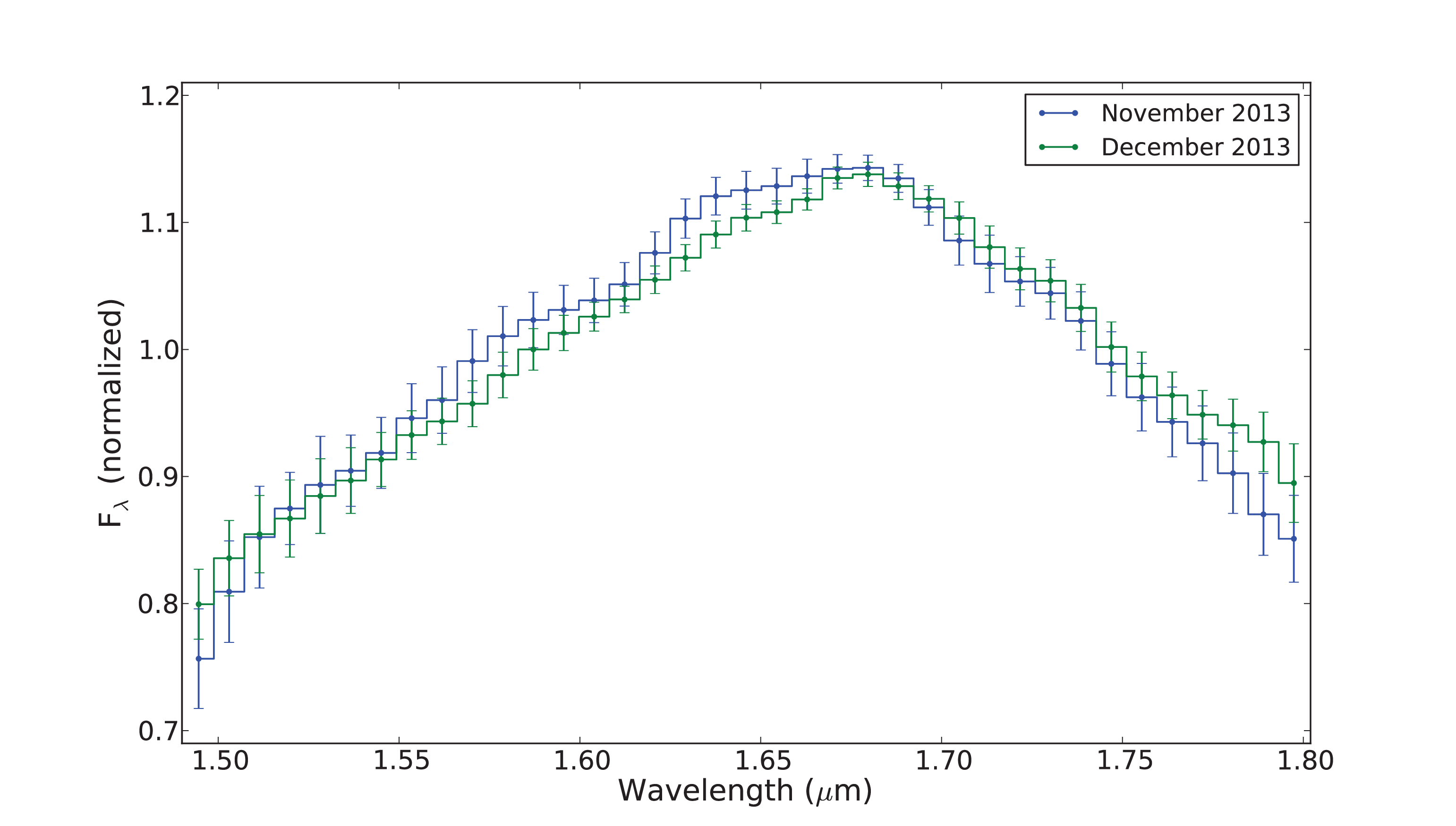}
\caption{$H$-band spectra of $\beta$ Pic b using both November and December 2013 observations from GPI. Both spectra are in agreement. The spectra were taken at different phases of the GPI commissioning process resulting in different effects on the light in the halo and PSF shape.}\label{Fig:Nov_Dec_Spectrum}
\end{figure}

We determined the spectrum using the flux of the PSF component of the background subtraction technique of the spline fit $+$ PSF to measure the flux from the injected 
reference PSF. This produces a measurements of the planet's flux in each spectral channel. Each of the individual spectra measured from the individual frames is independently normalized and combined together (Figure ~\ref{Fig:Nov_Dec_Spectrum}). To estimate the systematic errors and residuals, PSFs were generated from the satellite spots, injected with a flat spectrum at an identical radius from the host star into the individual frames, and then reduced in an identical manner.
Given that this is one of the first extracted
spectrum from the new GPI instrument, and that the halo of the star has significant color variation, it is possible that the overall spectral slope has an uncertainty of approximately 10$\%$.

\section{Results \& Discussion}\label{sec:results}

The spectrum discussed above has an SNR (per
wavelength channel) that matches or exceeds most previous broad band 
photometry. With this spectrum, we can estimate surface gravity and 
effective temperature as well as search for molecular absorption 
features and 
departures from stellar abundances.

The $H$-band spectrum has a clear peak at 1.68 $\mu$m defined by absorption on
either side. The location of this peak and the slopes on either side are
consistent with water absorption frequently seen in brown dwarf spectra. Based on 
previous photometric estimates of the effective temperature (1600--1700~K),
the primary opacity sources across the near-infrared are water,
collision-induced absorption (CIA) from H$_2$, and dust. There is no evidence
for additional molecular absorption (e.g., from methane or ammonia). The
$H$-band spectrum has a very triangular shape, a hallmark of low surface gravity
and further evidence of $\beta$ Pic b's low mass and youth.

The GPI $H$-band spectrum and existing ground-based photometry were compared to 
the model grids described in \citet{Barman2011a, Barman2011b}. An effective 
temperature of 1650$\pm 50$~K was found to best match these spectral data, in excellent 
agreement with previous photometric studies \citep{Bonnefoy2013, Currie2013, Males2014}.  
The best matching model is shown in Figure \ref{Fig:Spectrum3} and it agrees 
well with the visible to IR photometry. Broad-band photometric colors, however, 
are only modestly sensitive to surface gravity, emphasizing the need for spectral 
data.  Our $H$-band spectrum, as previously discussed, has an triangular shape that 
sensitively depends on surface gravity.  Our best matching models have 
$\log(g) = 4.0 \pm 0.25$ (cgs units) that, when taken into consideration along 
with the effective temperature of 1650~K, is consistent with evolutionary models 
between 10 and 20 Myrs for masses between 10 and 12 M$_{\rm Jup}$ 
\citep{Burrows1997,Chabrier2000}.

\begin{figure}
\epsscale{1.0}
\plotone{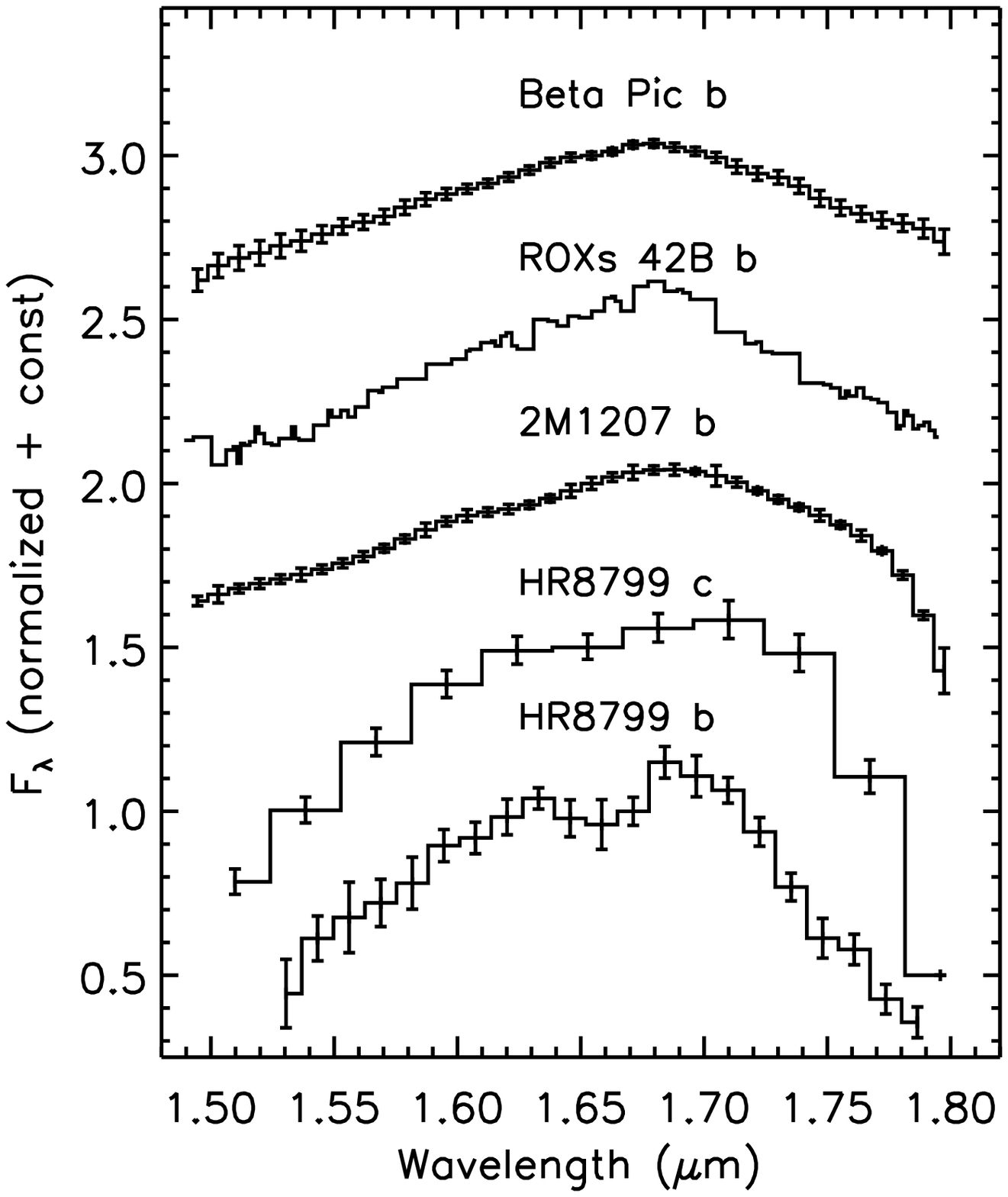}
\caption{$H$-band spectra of young, directly imaged planets.  The December 2013 Gemini Planet Imager spectrum of $\beta$ Pic b is plotted above the spectra of ROX 42b b \citep{Bowler2014}, 2M1207b \citep{Patience2010}, HR8799 c \citep{Oppenheimer2013} and HR8799 b \citep{Barman2011a}. Each of these objects is cooler then $\beta$ Pic b. Despite very different temperatures, ROXs 42B b, 2M1207b and \mbox{$\beta$ Pic b} have atmospheres with similar dominant opacity sources.  The differences between \mbox{$\beta$ Pic b} and HR8799 b and c highlights the spectral evolution of low gravity objects.}\label{Fig:Spectrum}
\end{figure}

\begin{figure}
\epsscale{1.0}
\plotone{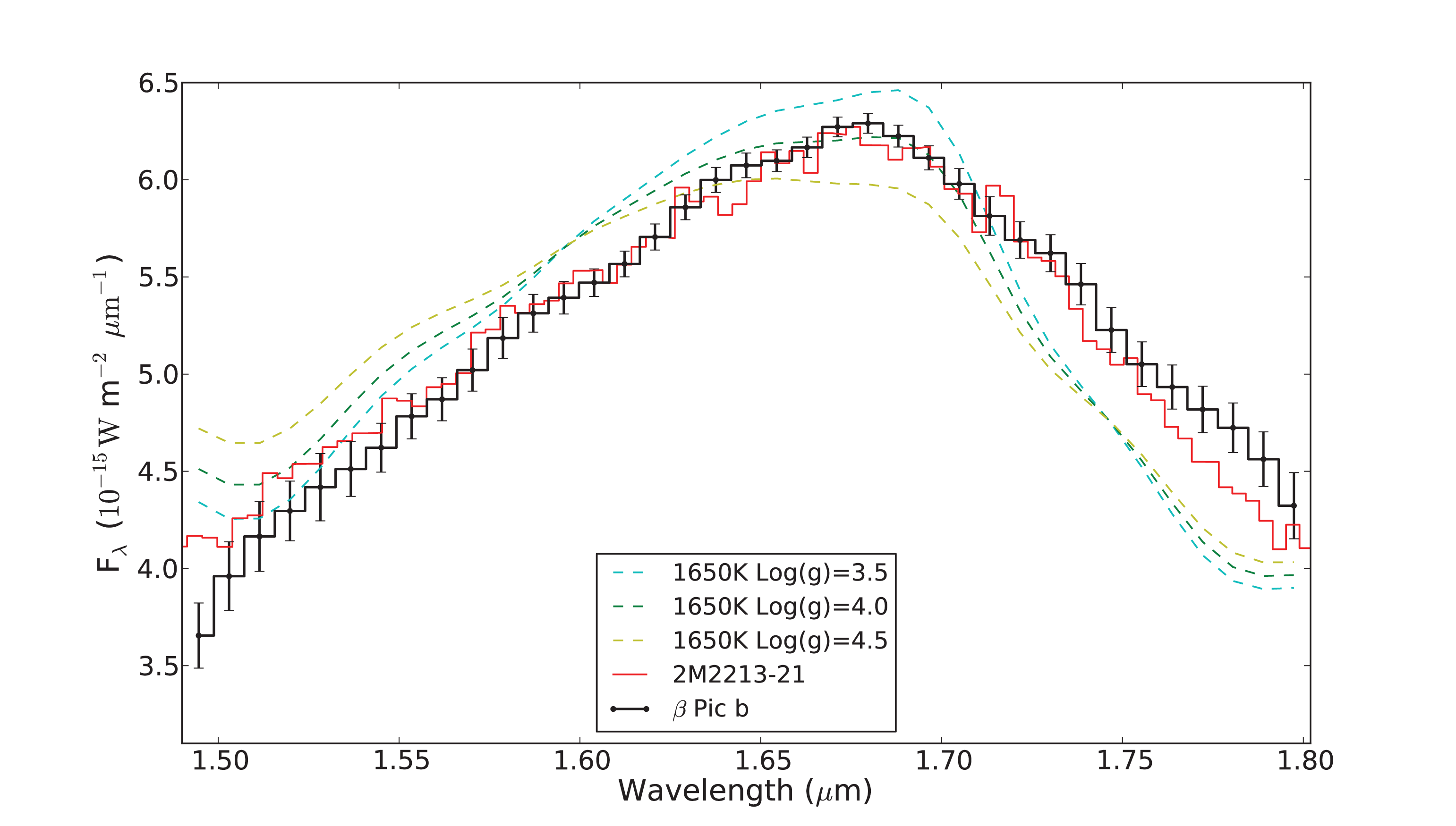}
\caption{The comparison of the $H$-band spectrum (black) to a 1650K model with 3 different gravities. All three models do not provide a perfect match to the spectrum. The $\log(g)=4.0$ model (green) comparison has the best fit but is offset from the observations by a constant slope. The young, low-gravity brown dwarf 2M2213-21 (red) has a better match to the spectrum then all 3 models. The agreement between the GPI spectrum and that of known low-gravity brown dwarfs strongly suggests that our GPI spectrum is mostly free of chromatic systematic errors and the discrepancies with the synthetic spectra are most likely the result of imperfect modeling (e.g., treatment of dust clouds). The spectra are normalized to match the flux measured in \citet{Males2014}.}\label{Fig:Spectrum3}
\end{figure}

Figure \ref{Fig:Spectrum} compares the December 2013 $\beta$ Pic b spectrum to those of 
other directly imaged planetary-mass companions: ROXs 42B b \citep{Bowler2014}, 
2M1207B \citep{Patience2010}, HR8799 b \citep{Barman2011a} and HR8799 c 
\citep{Oppenheimer2013}.  All of these objects are reported to exhibit low gravity. 
ROXs 42B b has a similar $H$-band spectrum as $\beta$ Pic b, though the former 
has a slightly steeper spectrum on either side of the peak, consistent with 
ROXs 42B b being slightly younger (5--10~Myr) or lower mass.  The other three 
planets shown in Figure \ref{Fig:Spectrum} are all cooler than $\beta$ Pic b 
by $\sim$ 500K to 800K.  Despite this large temperature difference, 
2M1207b and $\beta$ Pic b have similar $H$-band spectra. 2M1207b's $H$-band 
spectrum is shaped by a combination of low gravity, opacity from thick dusty 
clouds, and non-equilibrium chemistry that favors CO over methane \citep{Barman2011b}.
Non-equilibrium chemistry is less important in hotter objects like $\beta$ Pic
b that will have large CO/CH$_4$ ratio, regardless of vertical mixing.  Consequently,
despite very different temperatures, ROXs 42B b, 2M1207b and $\beta$ Pic b have
atmospheres with similar dominant opacity sources.  The $H$-band similarities
between these objects supports the idea that $\beta$ Pic b is low gravity (and
hence low mass) and 1 to 2 pressure scale heights near the photosphere.  
The differences between $\beta$ Pic
b and HR8799 b and c seen in Figure \ref{Fig:Spectrum} highlight the spectral 
evolution of low gravity objects from high to low effective temperatures.

The model spectra  (Figure \ref{Fig:Spectrum3}), however, do not match the $H$-band 
spectrum particularly well.  The best matching model under predicts the fluxes at
$\lambda >$ 1.7~$\mu$m while slightly over predicting the fluxes on the blue side
of the $H$-band peak. The net effect is a systematic tilt of 5 to 10\%\ between
the model and the data. Though a spectral offset of this magnitude may be present 
in the data, we found that most $H$-band spectra from a low gravity
brown dwarf spectral sequence \citep{Allers2013} agree extremely well with our GPI
spectrum. The best matching brown dwarf, 2M2213-21, has a reduced $\chi^2 = 1.7$
(see Fig. \ref{Fig:Spectrum3}) and the red-optical through $K$-band spectrum 
of 2M2213-21 also closely follows the $\beta$ Pic b photometry.
Like $\beta$ Pic b, 2M2213-21 is a young object with low gravity features 
and is possibly a member of 
the $\beta$ Pic moving group, at the $\sim 30\%$ level \citep{Manjavacas2014}.
The agreement between the GPI spectrum and that of known low-gravity brown dwarfs
strongly suggests that our GPI spectrum is mostly free of chromatic systematic 
errors and the discrepancies with the synthetic spectra are most likely the 
result of imperfect modeling (e.g., treatment of dust clouds). Such a systematic
discrepancy in the model spectra could bias the derived surface gravity, but it is unclear
by how much.  Allowing for a slight, $\pm 10$\%\ tilt in the model $H$-band
spectra yields much improved fits, but does not noticeably change the resulting
surface gravity.

\section{Conclusion}\label{sec:Discussion}

We present the first $H$-band spectrum of the extrasolar planet $\beta$ Pic b from the recently commissioned Gemini Planet Imager --- located on the Gemini South telescope --- which began commissioning in the Fall of 2013. The Gemini Planet Imager is a facility class instrument built to directly image and spectroscopically characterize young, Jupiter sized, self-luminous, extrasolar planets. We find that the spectrum of \mbox{$\beta$ Pic b} provides a new and insightful look at the atmospheres of these high-temperature low-gravity objects. While the best matching model does not perfectly match the $H$-band spectrum, the spectrum is remarkably similar to the young, low gravity brown dwarf 2M2213-21. We thus conclude that error most likely is derived from imperfect modeling of the atmosphere. With so few directly imaged planet spectra, the other known objects are estimated to be cooler then \mbox{$\beta$ Pic b}, and have a slightly different spectral shape. 

Currently, and in the near future, several extreme-AO instruments will be on-line with the capability to directly image the spectra of the extrasolar planets they find. While our $\beta$ Pic b data only cover the $H$-band, GPI is designed to measure spectra from $0.95-2.4\mu$m at a similar capability as our $H$-band data. These spectra will further our understanding of these high temperature low-gravity objects. The low resolution but great sensitivity of GPI is well designed to identify and characterize low gravity young exoplanets, as is demonstrated in our \mbox{$\beta$ Pic b} spectrum.

\acknowledgments

The authors would like to acknowledge the financial support of the Gemini Observatory, the NSF Center for Adaptive Optics at UC Santa Cruz, the NSF (AST-0909188; AST-1211562, AST-1405505), NASA Origins (NNX11AD21G; NNX10AH31G, NNX14AC21G), the University of California Office of the President (LFRP-118057), and the Dunlap Institute, University of Toronto. Portions of this work were performed under the auspices of the U.S. Department of Energy by Lawrence Livermore National Laboratory under Contract DE-AC52-07NA27344 and under contract with the California Institute of Technology/Jet Propulsion Laboratory funded by NASA through the Sagan Fellowship Program executed by the NASA Exoplanet Science Institute. We are indebted to the international team of engineers and scientists who worked to make GPI a reality.

{\it Facilities:} \facility{Gemini South (GPI)}.

%% ------------------------- END ACKNOWLEDGEMENTS ------------------------- %%

\begin{figure}
\epsscale{1.0}
\plotone{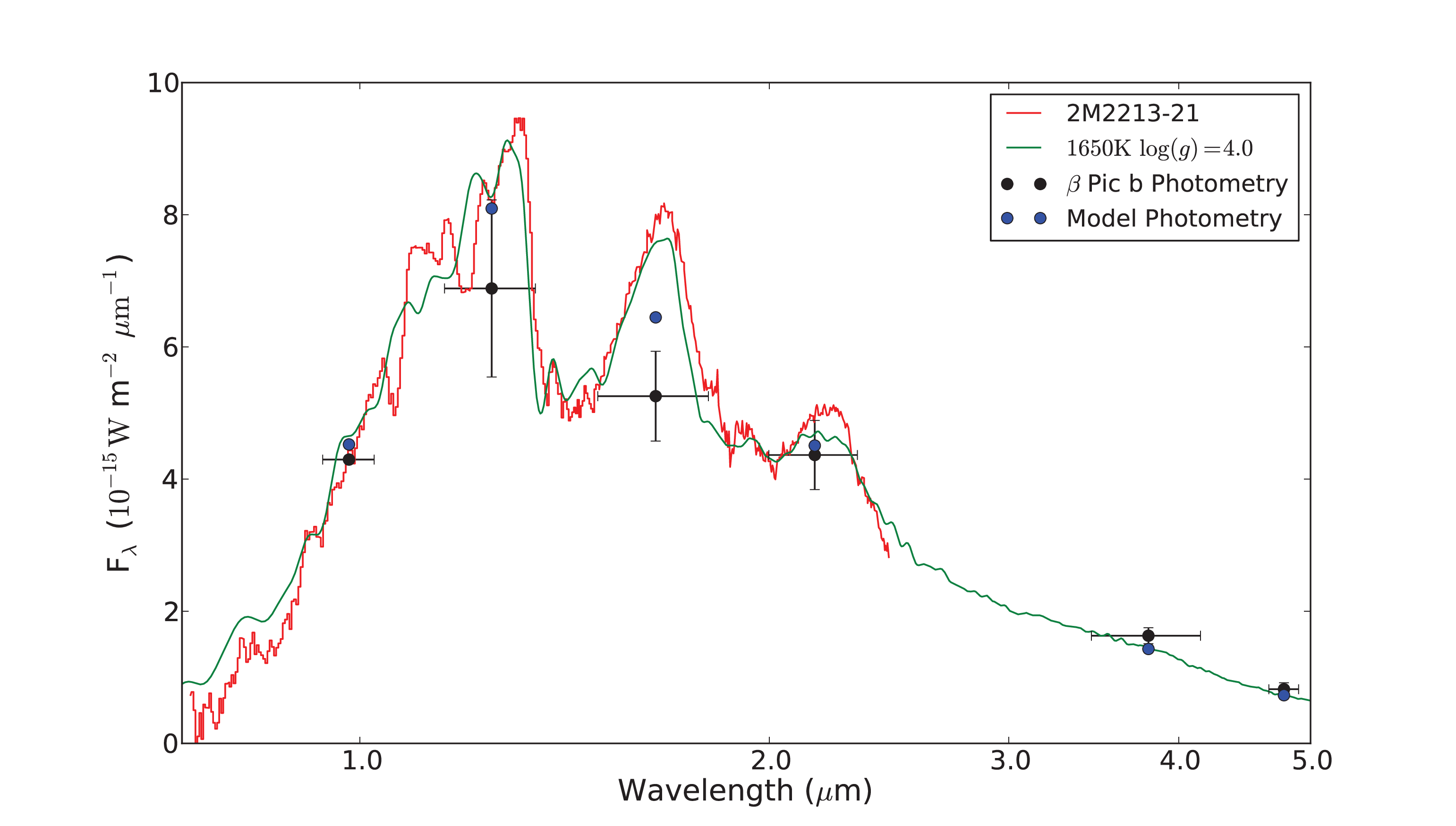}
\caption{We compare the model 1650K $\log(g)=4.0$ spectrum (green), and its predicted photometric points (blue) to the spectrum of 2M2213-21 (red) and the measured photometric points of $\beta$ Pic b (black) \citep{Males2014}.}\label{Fig:Spectrum2}
\end{figure}

\end{document}